\documentclass[%
 reprint,
superscriptaddress,
 amsmath,amssymb,
 aps,
prb,
]{revtex4-2}

\usepackage{graphicx}
\usepackage{subfigure}
\usepackage{float}
\usepackage{dcolumn}
\usepackage{bm}
\usepackage{rotating}
\usepackage{color}
\usepackage{ulem}
\usepackage{amsmath}
\usepackage{amsfonts}
\usepackage{amssymb}
\newcommand{\func}[1]{\mathrm{#1}}

\usepackage[colorlinks, citecolor=blue,urlcolor=blue,linkcolor=blue]{hyperref}
\usepackage{etoolbox}
\makeatletter
\let\oldbib@parse\bib@parse
\def\bib@parse#1{\oldbib@parse{#1}\def\BR@c@color{\color{red}}}
\makeatother

\begin{document}

\preprint{APS/123-QED}

\title{Scaling behavior of dissipative systems with imaginary gap closing} 

\author{Jinghui Pi}
\email{pijh14@gmail.com}
 \affiliation{The Chinese University of Hong Kong Shenzhen Research Institute, 518057 Shenzhen, China}
 \affiliation{Department of Physics, The Chinese University of Hong Kong, Shatin, New Territories, Hong Kong, China
} 

\author{Xingli Li}
\affiliation{Department of Physics, The Chinese University of Hong Kong, Shatin, New Territories, Hong Kong, China
} 
\affiliation{The Chinese University of Hong Kong Shenzhen Research Institute, 518057 Shenzhen, China}

\author{Yangqian Yan}
\email{yqyan@cuhk.edu.hk}
\affiliation{Department of Physics, The Chinese University of Hong Kong, Shatin, New Territories, Hong Kong, China
} 
\affiliation{State Key Laboratory of Quantum Information Technologies and Materials, The Chinese University of Hong Kong, Hong Kong SAR, China}
\affiliation{The Chinese University of Hong Kong Shenzhen Research Institute, 518057 Shenzhen, China}

 %



\begin{abstract}
Point-gap topology, characterized by spectral winding numbers, is crucial to non-Hermitian topological phases and dramatically alters real-time dynamics. In this paper, we study the evolution of quantum particles in dissipative systems with imaginary gap closing, using the saddle-point approximation method. For trivial point-gap systems, imaginary gap-closing points can also be saddle points. This leads to a single power-law decay of the local Green's function, with the asymptotic scaling behavior determined by the order of these saddle points. In contrast, for nontrivial point-gap systems, imaginary gap-closing points do not coincide with  saddle points in general. This results in a dynamical behavior characterized by two different scaling laws for distinct time regimes. In the short-time regime, the local Green's function is governed by the dominant saddle points and exhibits an asymptotic exponential decay. In the long-time regime, however, the dynamics is controlled by imaginary gap-closing points, leading to a power-law decay envelope.  Our findings advance the understanding of quantum dynamics in dissipative systems and provide predictions testable in future experiments.
\end{abstract}

\maketitle


\section{Introduction}

Non-Hermitian physics has achieved substantial progress over the past two decades \cite{Bender_2007, RevModPhys.87.61, RevModPhys.88.035002,el2018non,ashida2020non,RevModPhys.93.015005,doi:10.1146/annurev-conmatphys-040521-033133}. For instance, non-Hermitian Hamiltonians describe optical systems with gain and loss \cite{ruter2010observation, PhysRevLett.106.093902, PhysRevLett.106.093902,regensburger2012parity, peng2014parity, Feng_2017, Longhi_2017}, dissipative wave dynamics in classical networks \cite{PhysRevB.99.201411,PhysRevResearch.2.023265,helbig2020generalized,PhysRevLett.124.046401,zhang2021observation,zou2021observation,doi:10.1126/sciadv.adf7299,PhysRevLett.131.207201,LIU20241228,10.21468/SciPostPhys.16.1.002,ZHANG2025100102}, and quasiparticle dynamics in condensed matter systems \cite{Yoshida_2018, Shen_2018, Yamamoto_2019,PhysRevLett.125.227204,PhysRevB.107.195149,zheng2024experimental,PhysRevB.109.235139}. A prominent feature of such lattice systems is the non-Hermitian skin effect (NHSE) \cite{PhysRevLett.121.086803,PhysRevLett.121.136802,PhysRevB.97.121401,PhysRevB.99.201103,PhysRevLett.123.016805,PhysRevB.102.205118}, which refers to the localization of an extensive number of eigenstates at the edges under the open boundary condition (OBC). This anomalous boundary localization phenomenon is closely related to the point-gap topology of non-Hermitian Bloch Hamiltonians \cite{PhysRevX.9.041015,PhysRevLett.124.086801,PhysRevLett.132.136401,PhysRevLett.124.056802}, causing a high sensitivity of the spectrum to boundary conditions and reshaping the conventional bulk-boundary correspondence \cite{helbig2020generalized,Xiao2020Non,doi:10.1073/pnas.2010580117,PhysRevLett.123.246801,PhysRevLett.123.066404,PhysRevLett.125.126402,PhysRevLett.127.116801,li2021quantized,Zhang2022Universal, PhysRevX.14.021011,xc42-7hcg}. Another unique feature of dissipative quantum systems is the closing of the Liouvillian gap (or imaginary gap) \cite{PhysRevA.86.012116,PhysRevA.86.013606, PhysRevA.87.012108,PhysRevA.98.042118}, which results in algebraic damping behavior and a divergence of the relaxation time \cite{PhysRevLett.111.150403, PhysRevA.90.033612, PhysRevE.92.042143, PhysRevLett.123.170401,PhysRevLett.127.070402,PhysRevResearch.2.043167,PhysRevLett.125.230604,PhysRevResearch.4.023160,PhysRevB.109.214311}.
   
Notably, boundary effects also induce distinctive non-Hermitian dynamical phenomena, including unidirectional amplification \cite{PhysRevX.8.041031,wanjura2020topological,PhysRevB.103.L241408}, enhanced quantum sensing \cite{lau2018fundamental,mcdonald2020exponentially,PhysRevLett.125.180403, PhysRevResearch.4.013113}, and novel dynamics of entanglement \cite{PhysRevResearch.2.033017,PhysRevLett.126.170503,PhysRevB.104.L161107,PhysRevB.107.L020403, PhysRevX.13.021007} and wave packet propagation \cite{PhysRevLett.128.157601,PhysRevLett.131.036402,PhysRevLett.128.120401,PhysRevLett.133.070801,PhysRevA.109.022236}. During time evolution, the influence of boundary terms diminishes with increasing system size, rendering non-Hermitian dynamics independent of boundary conditions in the thermodynamic limit \cite{PhysRevB.104.125435}.  This behavior is intuitive for trivial point-gap systems, as their energy spectra are also independent of boundary conditions in this limit. In contrast, for nontrivial point-gap systems, energy spectra under OBC differ significantly from those under periodic boundary conditions (PBC) \cite{PhysRevX.9.041015,PhysRevLett.124.086801,PhysRevLett.132.136401,PhysRevLett.124.056802}. For finite-sized systems, it is unclear how to detect this spectral discrepancy and incorporate boundary effects through a real-time dynamical process. Moreover, for systems with imaginary gap closing, the distinction in bulk dynamics between trivial and nontrivial point-gap phases also remains largely unexplored.

In this work, we investigate the long-time dynamics of quantum particles initially released at bulk sites of dissipative systems with imaginary gap closing. To facilitate the analysis of these systems, we introduce the non-Hermitian multi-band Green's function and the concept of point-gap topology. For systems with a trivial point gap, the imaginary gap-closing points can also be saddle points of the non-Hermitian Bloch Hamiltonian.  In such cases, the local Green's function exhibits a power-law decay, scaling asymptotically as $t^{-1/n}$, where $n$ is the order of the dominant saddle points. In contrast, for systems with a nontrivial point gap, the imaginary gap-closing points do not coincide with saddle points. As a result, the local Green's function demonstrates two distinct dynamical regimes. In the short-time regime, the evolution is governed by the dominant saddle points and shows an asymptotic exponential decay. Beyond this transient period, the influence of the boundary becomes significant. The subsequent dynamics are then controlled by the imaginary gap-closing points, and the local Green's function exhibits a power-law evolution. This long-time scaling is determined by the saddle point of the world-line Green's function, which is related to the group velocities at the corresponding imaginary gap-closing points. The aforementioned asymptotic scaling behavior in imaginary gap-closing systems is derived theoretically using the saddle-point approximation and is confirmed numerically.
Our findings enhance the understanding of quantum dynamics in dissipative systems and provide a theoretical foundation for future experimental observation of saddle-point dynamics in non-Hermitian systems.

The remainder of this paper is organized as follows. In Sec.~\ref{sec2}, we present the general form of the non-Hermitian Green's function in the time domain and a brief introduction to point-gap topology. The long-time dynamics of imaginary gap-closing systems are then analyzed for two cases, based on the saddle point approximation of the Green's function. In Sec.~\ref{sec3}, we study the scaling behavior in systems with trivial point-gap topology using a concrete dissipative two-band model. Sec.~\ref{sec4} extends this analysis to systems with a nontrivial point-gap topology via an associated two-band model. Finally, a summary of the main conclusions and an outlook for future work are provided in Sec.~\ref{sec5}.

\section{Multi-band real-time Green functions and Point-gap topology}\label{sec2}

For a generic one-dimensional (1D) multiband non-Hermitian tight-binding
system, the real-space Hamiltonian reads:%
\begin{equation}
 H=\sum_{x,y}\sum_{a,b}^{q}t_{y-x}^{ab}\left\vert x,a\right\rangle
\left\langle y,b\right\vert ,   
\end{equation}
where $x$ and $y$ denote the spatial locations of the unit cells, and $%
a,b=1,2,\ldots ,q$ represent the internal orbitals within each unit cell.
The hopping amplitude $t_{y-x}^{ab}$ has a finite range and vanishes for $%
\left\vert y-x\right\vert >N$. The system's evolution can be captured by the
single-particle Green's function in the time domain, which is defined as%
\begin{equation}
G_{ab}\left( x,y; t\right) =\left\langle x,a\right\vert e^{-iHt}\left\vert
y,b\right\rangle .
\label{eq2}
\end{equation}
By imposing periodic boundary conditions and performing a Fourier transformation, the Green's function (\ref{eq2}) in the thermodynamic limit can be
recast as an integral over the Brillouin zone (BZ):%
\begin{equation}
\begin{aligned}
 G_{ab} \left( x,y;t\right) =&\int_{0}^{2\pi }\frac{dk}{2\pi }e^{ik\left(
x-y\right) }\left\langle a\right\vert e^{-ih\left( k\right) t}\left\vert
b\right\rangle  \\
=& \oint\nolimits _{\left\vert \beta \right\vert =1} \frac{d\beta }{2\pi
i\beta }\beta ^{x-y}\left\langle a\right\vert e^{-ih\left( \beta \right)
t}\left\vert b\right\rangle,
\label{Eq3}
\end{aligned}  
\end{equation}
where $\beta =e^{ik}$ is the Bloch phase factor and $h\left( \beta \right)
_{ab}=\sum_{l=-N}^{N}t_{l}^{ab}\beta ^{l}$ is the non-Hermitian Bloch
Hamiltonian. To calculate the integral (\ref{Eq3}), we can diagonalize $h(\beta)$ using its left and right eigenstates. As a result, the
matrix element of the time evolution operator $e^{-ih\left( \beta \right) t}$
becomes
\begin{equation}
\left\langle a\right\vert e^{-ih\left( \beta \right) t}\left\vert
b\right\rangle =\sum_{n=1}^{q}e^{-iE_{n}\left( \beta \right) t}R_{n,a}\left(
\beta \right) L_{n,b}\left( \beta \right) ,
\end{equation}
where $R_{n,a}(\beta) = \langle a | R_{n}(\beta) \rangle$, $L_{n,b}(\beta) = \langle L_{n}(\beta) | b \rangle$, and $E_{n}(\beta)$ is the $n$-th band eigenvalue of $h(\beta)$. Here, $\langle L_{n}(\beta) |$ and $| R_{n}(\beta) \rangle$ denote the corresponding left and right eigenstates.
By defining $g_{n}^{ab}\left( \beta \right)
=R_{n,a}\left( \beta \right) L_{n,b}\left( \beta \right) $, we obtain a
compact representation of the Green's function:%
\begin{equation}
G_{ab}\left( x,y;t\right) =\sum_{n}  \oint\nolimits_{\left\vert \beta
\right\vert =1}\frac{d\beta }{2\pi i\beta }\beta ^{x-y}g_{n}^{ab}\left(
\beta \right) e^{-iE_{n}\left( \beta \right) t}.
\end{equation}
For a system with onsite dissipation, the energy spectrum is bounded by the
condition $ \func{Im}\left[ E_{n}\left( \beta \right) \right] \leq 0$.
Especially, if points $\beta _{0}$ exist in the BZ such that $\func{Im}\left[
E_{n}\left( \beta_{0} \right) \right] =0$ for some band $E_n$, the imaginary
(dissipative) gap of the system is closed. We focus on this case in the
remainder of this paper.

An important concept in non-Hermitian topological phases
is the point gap, which is characterized by a winding number of the spectrum
in the complex energy plane. Specifically, for a reference point $E_{b}\in 
\mathbb{C}$, the following winding number can be defined over the BZ \cite{PhysRevX.9.041015,PhysRevLett.124.086801,PhysRevLett.125.126402}:%
\begin{equation}
W\left( E_{b}\right)  =\oint\nolimits_{\left\vert \beta
\right\vert =1}\frac{d\beta
}{2\pi i}\partial_{\beta}\ln \det[h\left(  \beta\right)  -E_{b}].
\label{W(E_b)}
\end{equation}
If $W\left( E_{b}\right) =0$ for all $E_{b}\in \mathbb{C}$,  the point-gap
of the system is trivial. In this case, the spectrum under PBC collapses
onto the spectrum under OBC in the thermodynamic limit. In contrast,  if $%
W\left( E_{b}\right) \neq 0$ for a given $E_{b}$, the system possesses a
nontrivial point gap, implying that the PBC spectrum encloses a finite area
in the complex energy plane. Consequently, the corresponding OBC spectrum is
enclosed by the PBC spectrum, and the eigenstates feature the NHSE \cite{PhysRevLett.124.086801,PhysRevLett.125.126402}. To
obtain the OBC spectra from the non-Hermitian Bloch Hamiltonian $h\left(
\beta \right) $, one can use the concept of the generalized Brillouin zone
(GBZ). For a given energy $E$, the characteristic equation $f\left( \beta
,E\right) =\det \left[ h\left( \beta \right) -E\right] =0$ yields $2M=2qN$
solutions. When these solutions are ordered by their magnitude, $\left\vert
\beta _{1}\left( E\right) \right\vert \leq \left\vert \beta _{2}\left(
E\right) \right\vert \leq \cdots \leq \left\vert \beta _{2M}\right\vert $,
the GBZ is determined by the condition $\left\vert \beta _{M}\left( E\right)
\right\vert =\left\vert \beta _{M+1}\left( E\right) \right\vert $ \cite{PhysRevLett.121.086803,PhysRevLett.123.066404,PhysRevLett.125.126402}, which traces a
closed curve in the complex $\beta $-plane.

\section{Scaling behavior of imaginary gap-closing systems with a trivial point-gap topology} \label{sec3}

\begin{figure}[t]
    \includegraphics[width=\columnwidth]{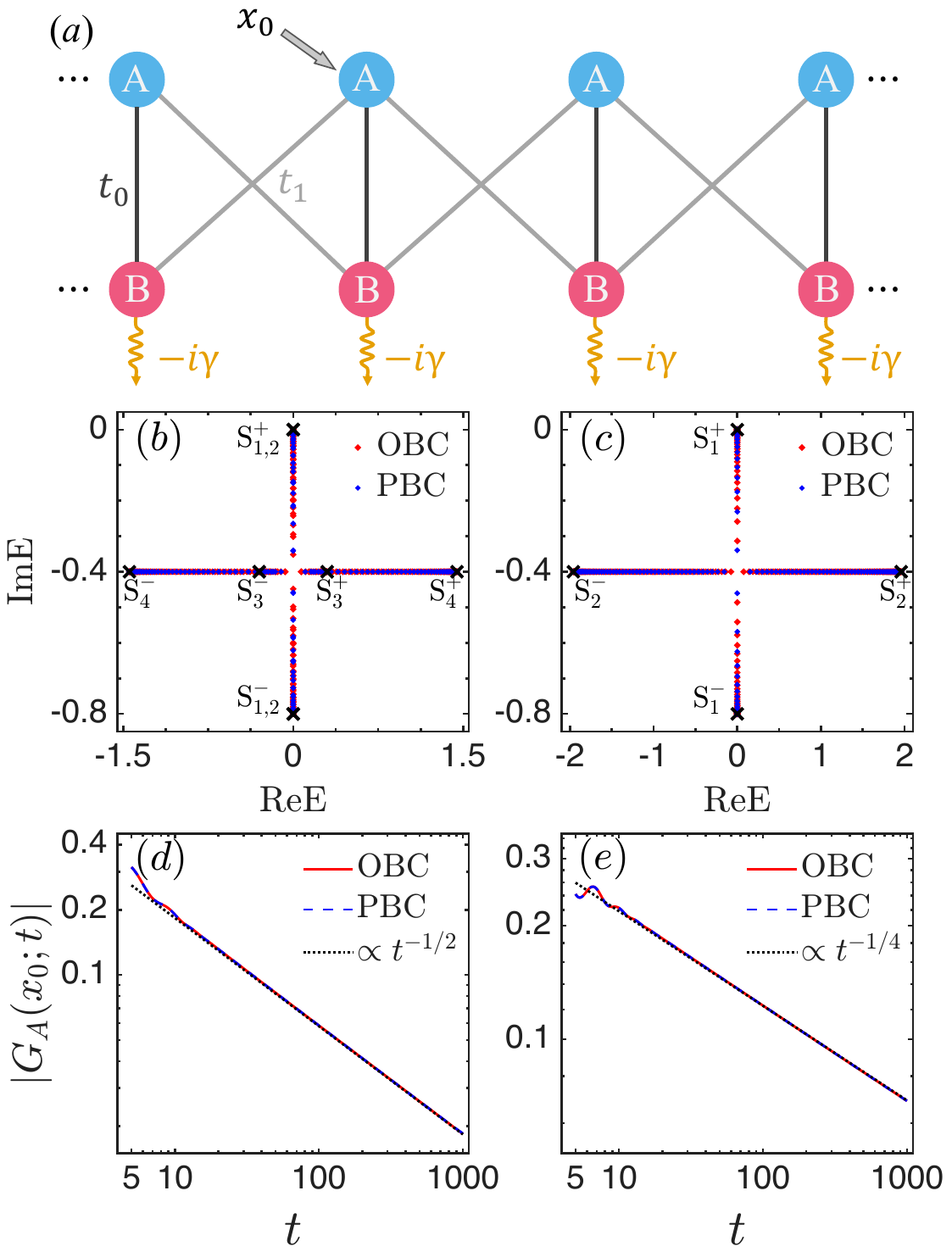}
  \caption{ (a) Schematic of a one-dimensional lossy ladder lattice with trivial point-gap topology, which corresponds to the Hamiltonian in Eq. (\ref{III(1)}). Each unit cell, labeled by the spatial coordinate $x$, comprises two sites, denoted $A$ and $B$. The particle is initially localized on the $A$ sublattice of the unit cell at $x_0$.  The energy spectra and corresponding numerical results for the local Green's function  $G_{A}\left(  x_{0};t\right)=\left\langle x_{0},A\left\vert e^{-iHt}\right\vert
x_{0},A\right\rangle $ are as follows:
   (b)-(c) OBC (red points) and PBC (blue points) energy spectra $E_{\pm}$, with saddle points $S^{\pm}_i$ explicitly marked by cross symbols. The saddle points $S^{+}_{1}$ and $S^{+}_{2}$ in (b), as well as the threefold saddle point $S^{+}_{1}$ in (c), are also imaginary gap-closing points. (d)-(e) Numerical results for the scaling of $|G_{A}\left(  x_{0};t\right)|$ with bulk $x_0=75$ under OBC (red solid lines) and PBC (blue dashed lines); these agree well with the theoretical results (black dashed lines). In (b) and (d), the parameters are $t_0=0.5$,$t_1=0.5$, and $\gamma=0.8$; in (c) and (e), $t_0=1.0$, $t_1=0.5$, and $\gamma=0.8$. The number of unit cells is $L=150$ for the numerical calculation. } 
  \label{Fig1}
\end{figure}

In this section, we discuss imaginary gap-closing systems with a trivial point gap. For concreteness, we consider a one-dimensional lossy lattice ladder, as shown
in Fig.~\ref{Fig1}(a). The tight-binding Hamiltonian reads:%
\begin{equation}
\begin{aligned}
H  =& \sum_{n}\left(  t_{0}c_{n,A}^{\dagger}c_{n,B}+t_{1}c_{n+1,A}^{\dagger
}c_{n,B}+t_{1}c_{n-1,A}^{\dagger}c_{n,B}\right)  \\
& +\text{H.c.} -\sum_{n}i\gamma c_{n,B}^{\dagger}c_{n,B},
\label{III(1)}
\end{aligned}
\end{equation}
where $t_0$ and $t_1$ are the intracell and intercell hopping amplitudes, respectively, $\gamma$ is the onsite dissipation on sublattice B, and H.c. denotes the Hermitian conjugate.
By applying a Fourier transformation and making the substitution $e^{ik}\rightarrow \beta $, we obtain the generalized non-Hermitian Bloch
Hamiltonian:%
\begin{equation}
h\left(  \beta\right)  =h_{x}\left(  \beta\right)  \sigma_{x}+\frac{i\gamma
}{2}\sigma_{z}-\frac{i\gamma}{2}I,
\label{III(2)}
\end{equation}
where $h_{x}\left(  \beta\right)  =t_{0}+t_{1}\left(  \beta+\beta^{-1}\right)
$. Here, $\sigma_{x,y,z}$ are the Pauli matrices, with sublattice A (B) corresponding to the up (down) component of a pseudo-spin, while $I$ denotes the
identity matrix. The energy spectrum of Eq.~(\ref{III(2)}) is
\begin{equation}
E_{\pm}\left(  \beta\right)  =-\frac{i\gamma}{2}\pm\sqrt{h_{x}^{2}\left(
\beta\right)  -\frac{\gamma^{2}}{4}},
\label{ES1}
\end{equation}
which consists of several open curves in the complex energy plane, as shown in
Figs.~\ref{Fig1}(b) and \ref{Fig1}(c). Hence, the point gap of the system is always
trivial; that is, the winding number $W\left(  E_{b}\right)  $ defined in
Eq.~(\ref{W(E_b)}) is equal to zero for any arbitrary complex reference energy
$E_{b}\in\mathbb{C}$. As a result, the energy spectra under both OBCs and PBCs
collapse to the same curves in the thermodynamic limit, implying the absence
of the NHSE.  Due to the dissipative nature of the system, the imaginary part of the energy must be less than or equal to zero, that is, $\operatorname{Im}E_{\pm}\left(
\beta\right)  \leq0$. In particular, when $h_{x}\left(  \beta\right)  =0$, the
Bloch energy spectrum (\ref{ES1}) touches the real axis (i.e., $\max\left[
\operatorname{Im}E_{+}\left(  \beta\right) \right]  =0 $), thus closing
the imaginary gap \cite{PhysRevB.109.214311}. Given that $\left\vert \cos k\right\vert \leq1$, this
condition is satisfied only when $\left\vert t_{0}\right\vert \leq\left\vert
2t_{1}\right\vert $. 

On the other hand, at the saddle points of the Bloch
Hamiltonian (\ref{III(2)}), we have $\frac{dE_{\pm}\left(  \beta\right)  }{d\beta}%
|_{\beta=\beta_{s}}=0$, which requires%
\begin{equation}
h_{x}\left(  \beta_{s}\right)  \left(  1-\beta_{s}^{-2}\right)  =0.
\label{sdeq1}
\end{equation}
Thus, in addition to the solutions $\beta_{s}=\pm1$, any other saddle points
should satisfy $h_{x}\left(  \beta_{s}\right)  =0$. This implies that the system's imaginary gap-closing points are also its saddle points. However, the
converse is not necessarily true. This distinction arises because
imaginary gap-closing points are restricted to the unit circle $\left\vert
\beta\right\vert =1$, whereas saddle points can exist anywhere in the complex
$\beta$ plane. For instance, when $\left\vert t_{0}\right\vert >\left\vert
2t_{1}\right\vert $, the equation $h_{x}\left(  \beta_{s}\right)  =0$  yields
two roots with $\left\vert \beta\right\vert \neq1$. In this scenario, the
system has saddle points but lacks imaginary gap-closing points.
Furthermore, the configuration of saddle points exhibits two distinct
behaviors depending on the system's parameters. For $\left\vert
t_{0}\right\vert <\left\vert 2t_{1}\right\vert $, equation $h_{x}\left(
\beta_{s}\right)  =0$ has two distinct solutions, $\beta_{s}=e^{\pm ik_{s}}$,
where $\cos k_{s}=-t_{0}/\left(  2t_{1}\right)  $. These solutions lie on the
unit circle and correspond to two imaginary gap-closing points at an energy of
$E_{+}\left(  \beta\right)  =0$, as shown in Fig.~\ref{Fig1}(b). In contrast, at the critical parameters $\left\vert t_{0}\right\vert =\left\vert 2t_{1}\right\vert $, the equation for saddle points (\ref{sdeq1}) has a threefold degenerate solution $\beta
_{s}=-1$. This point also corresponds to an imaginary gap closing at
$E_{+}\left(  \beta\right)  =0$, as shown in Fig.~\ref{Fig1}(c).

We now examine the dynamic behavior of a wave packet along the 1D
ladder governed by the Hamiltonian in Eq.~(\ref{III(1)}). The initial state is a delta
function localized at the non-decaying sublattice $A$ of the bulk $x_{0}$, such that
$\psi_{x}^{A}(0)=\delta_{x,x_{0}}$ and $\psi_{x}^{B}(0)=0$. In this dissipative
system, the amplitude at the initial site is defined as $\left\langle x_{0},A\left\vert e^{-iHt}\right\vert
x_{0},A\right\rangle $. This corresponds to the time-domain local Green's function
$G_{AA}\left(  x_{0},x_{0};t\right) $ (abbreviated as $G_{A}\left(  x_{0};t\right)$ for simplicity), which can be probed in experimental systems \cite{PhysRevA.111.053308}. As a result, in the thermodynamic limit (i.e., for a large system size), the local Green's function  $G_{A}\left(  x_{0};t\right)$ under PBC takes the form of
\begin{equation}
  G_{A}\left(  x_{0};t\right) =\sum_{n=\pm}%
{\displaystyle\oint\nolimits_{\left\vert \beta\right\vert =1}}
\frac{d\beta}{2\pi i\beta}g_{n}^{AA}\left(  \beta\right)  e^{-i E_{n}\left(
\beta\right)t },  
\label{11}
\end{equation}
where $g_{n}^{AA}\left(  \beta\right)  =\left\langle A|R_{n}\left(
\beta\right)  \right\rangle \left\langle L_{n}\left(  \beta\right)
|A\right\rangle $, with $\left\vert R_{n}\left(  \beta\right)  \right\rangle $
and $\left\langle L_{n}\left(  \beta\right)  \right\vert $ denoting the right
and left eigenstates of the Bloch Hamiltonian (\ref{III(2)}), respectively. To evaluate the long-time behavior of the contour integrals in Eq.~(\ref{11}), the steepest descent method is a commonly employed approach (see Appendix \ref{appB} for further details) \cite{bleistein1986asymptotic}. The
fundamental principle involves the deformation of the original integral path
$\left\vert \beta\right\vert =1$ into a new path $\mathcal{C}^{\prime}$ in the
complex $\beta$ plane. This deformation adheres to two key conditions: (i) the
contour $\mathcal{C}^{\prime}$ passes through a saddle point $\beta_{s}$,
where $\operatorname{Im}E_{n}\left(  \beta_{s}\right)  \geq\operatorname{Im}%
E_{n}\left(  \beta\right)  $ for all $\beta\in\mathcal{C}^{\prime}$; (ii) the
real part of $E_{n}\left(  \beta\right)  $ remains constant along the path
$\mathcal{C}^{\prime}$, namely $\operatorname{Re}E_{n}\left(  \beta\right)
=\operatorname{Re}E_{n}\left(  \beta_{s}\right)  $ for all $\beta
\in\mathcal{C}^{\prime}$. Hence, this new path must represent the steepest descent direction of $\operatorname{Im}%
E_{n}\left(  \beta\right)  $ from the saddle point $\beta_{s}$ (see more details in Appendix \ref{appA}). Then, for large $t$, it follows that the contour integral is
asymptotically dominated by this saddle point, yielding: 
\begin{equation}
%
{\displaystyle\oint\nolimits_{\left\vert \beta\right\vert =1}}
\frac{g_{n}^{AA}\left(  \beta\right)  }{2\pi i\beta}e^{-iE_{n}\left(
\beta\right)  t}d\beta\overset{t\rightarrow\infty}{\longrightarrow}\alpha
t^{-\frac{1}{2}}e^{-iE_{n}\left(  \beta_{s}\right)t},%
\end{equation}
where $\alpha=\frac{g_{n}^{AA}\left(  \beta_{s}\right)  }{\beta_{s}\sqrt{2\pi
i^{3}E_{n}^{\prime\prime}\left(  \beta_{s}\right)  }}$, provided that
$g_{n}^{AA}\left(  \beta_{s}\right)  \neq0$ and $E_{n}^{\prime\prime}\left(
\beta_{s}\right)  \equiv\frac{d^{2}E_{n}\left(  \beta\right)  }{d\beta^{2}%
}|_{\beta_{s}}\neq0$. 

When the imaginary gap-closing condition is satisfied, the contribution of the
$E_{+}\left(  \beta\right)  $ band to the contour integral in Eq.~(\ref{11}) is
dominated by the corresponding saddle point where $E_{+}\left(  \beta\right)
=0$. If $\left\vert t_{0}\right\vert <\left\vert 2t_{1}\right\vert $, we can
select a deformed contour $\mathcal{C}^{\prime}$ that passes through this
saddle point and apply the steepest descent method. This yields a power-law
decay scaling of $t^{-1/2}$ for the contour integral of the $E_{+}\left(
\beta\right)  $ band. Since $\operatorname{Im}E_{-}\left(  \beta\right)
\leq-\gamma/2$, the contribution of the $E_{-}\left(  \beta\right)  $ band
contour integral in Eq.~(\ref{11}) becomes exponentially small for large $t$ and
can, therefore, be neglected. Consequently, in the long time limit, the
local Green's function $G_{A}\left(  x_{0};t\right) $ exhibits the following
asymptotic power-law scaling behavior:%
\begin{equation}
\lim_{t\rightarrow\infty}G_{A}\left(  x_{0};t\right) \propto t^{-1/2},
\end{equation}
provided that $\left\vert t_{0}\right\vert <\left\vert 2t_{1}\right\vert $. We
numerically calculate $G_{A}\left(  x_{0};t\right) $ under both PBC
and OBC. In Fig. \ref{Fig1}(d), we show that the results for both conditions collapse onto nearly identical evolution curves and agree well with
the scaling of the theoretically predicted $t^{-1/2}$ power-law at large $t$.

However, the aforementioned steepest descent method fails when $\left\vert
t_{0}\right\vert =\left\vert 2t_{1}\right\vert $. At these critical
parameters, the imaginary gap-closing energy corresponds to a fourth-order saddle point at $\beta_{s}=-1$, leading to the vanishing of the
second-order derivative, $\frac{d^{2}E_{+}\left(  \beta\right)  }{d\beta^{2}%
}|_{\beta=-1}$. In this case, the long-time asymptotic behavior of the contour
integral $%
{\displaystyle\oint\nolimits_{\left\vert \beta\right\vert =1}}
\frac{g_{+}^{AA}\left(  \beta\right)  }{2\pi i\beta}e^{-iE_{+}\left(
\beta\right)  t}d\beta$ is still dominated by this fourth-order saddle point. To
evaluate this integral, we expand $E_{+}\left(  \beta\right)  $ near
$\beta_{s}=-1$. In the neighborhood of this point,  $E_{+}\left(
\beta\right)  \approx\frac{E_{+}^{\left(  4\right)  }\left(  -1\right)  }%
{4!}\left(  \beta-1\right)  ^{4}$, where $E_{+}^{\left(  4\right)  }\left(
-1\right)  =\frac{d^{4}E_{+}\left(  \beta\right)  }{d\beta^{4}}|_{\beta=-1}$.
Therefore, a modified steepest descent method is required. By deforming the
contour $\left\vert \beta\right\vert =1$ into the steepest descent path
$\mathcal{C}^{\prime}$ around the fourth-order saddle point $\beta_{s}=-1$, in
the large $t$ limit, the leading order contribution of the integral is given
by \cite{bleistein1986asymptotic}
\begin{equation}
\lim_{t\rightarrow\infty}%
{\displaystyle\oint\nolimits_{\left\vert \beta\right\vert =1}}
\frac{g_{+}^{AA}\left(  \beta\right)  }{2\pi i\beta}e^{-iE_{+}\left(
\beta\right)  t}d\beta\sim\alpha^{\prime}t^{-\frac{1}{4}},
\end{equation}
where $\alpha^{\prime}=-\frac{g_{+}^{AA}\left(  -1\right)  \Gamma\left(
1/4\right)  }{8\pi i\left[ -E_{+}^{\left(  4\right)  }\left(
-1\right)  /24\right]  ^{1/4}}$. As a result, the asymptotic scaling of the local Green's function
$G_{A}\left(  x_{0};t\right)    $ at these critical values $\left\vert
t_{0}\right\vert =\left\vert 2t_{1}\right\vert $ is%
\begin{equation}
\lim_{t\rightarrow\infty}G_{A}\left(  x_{0};t\right)   \propto t^{-1/4}.
\end{equation}
This theoretical prediction agrees closely with the numerical calculations in the long time limit, for both PBC
and OBC, as shown in Fig.~\ref{Fig1}(e). 

The aforementioned results can, in fact, be extended to a more general
dissipative model. Specifically, if the imaginary gap-closing point is also an $n$-th order saddle point $\beta_{s}$ of the energy band $E_{m}\left(
\beta\right)  $,  the local Green's function
$G_{A}\left(  x_{0};t\right)$ can be evaluated using a higher-order
saddle point approximation. An $n$-th order saddle point is defined by the
conditions $E_{m}^{\left(  p\right)  }\left(  \beta_{s}\right)  =0$ for
$p=1,2,\cdots,n-1,$ and $E_{m}^{\left(  n\right)  }\left(  \beta_{s}\right)
\neq0$, where $E_{m}^{\left(  p\right)  }\left(  \beta_{s}\right)
=(d^{p}E_{m} /d\beta^{p})|_{\beta_{s}}$.
Applying the approximation yields the following asymptotic expression \cite{bleistein1986asymptotic}:%
\begin{equation}
  G_{A}\left(  x_{0};t\right)  \sim 
\frac{g_{m}^{AA}\left(  \beta_{s}\right) \Gamma\left(  1/n\right)  }{n\left[  -E_{m}^{\left(  n\right)
}\left(  \beta_{s}\right)/(n!)\right]  ^{1/n}}t^{-1/n},  
\end{equation}
implying that the long time evolution of $G_{A}\left(  x_{0};t\right)  $
exhibits a characteristic scaling behavior of $t^{-1/n}$. 

The above discussion does not always hold for imaginary gap-closing systems with a trivial point-gap topology. In fact, at these imaginary gap-closing points $\beta_0$, we have $\operatorname{Im} \partial_{\beta}E(\beta)|_{\beta=\beta_0}=0$. Hence, if $\operatorname{Re}\partial_{\beta}E(\beta)|_{\beta=\beta_0}=0$, these points are also saddle points and correspond to the endpoints of the energy spectrum $E(\beta)$. However, it is also possible that $\operatorname{Re}\partial_{\beta}E(\beta)|_{\beta=\beta_0}\neq0$, which means that $\beta_0$ is not a saddle point of $E(\beta)$. In this case, the energy spectrum $E(\beta)$ is tangent to the $\operatorname{Re} E$ axis at $\beta_0$ in the complex plane.

\section{Scaling behavior of imaginary gap-closing systems with a nontrivial point-gap topology}\label{sec4}

\begin{figure}[t]
    \includegraphics[width=\columnwidth]{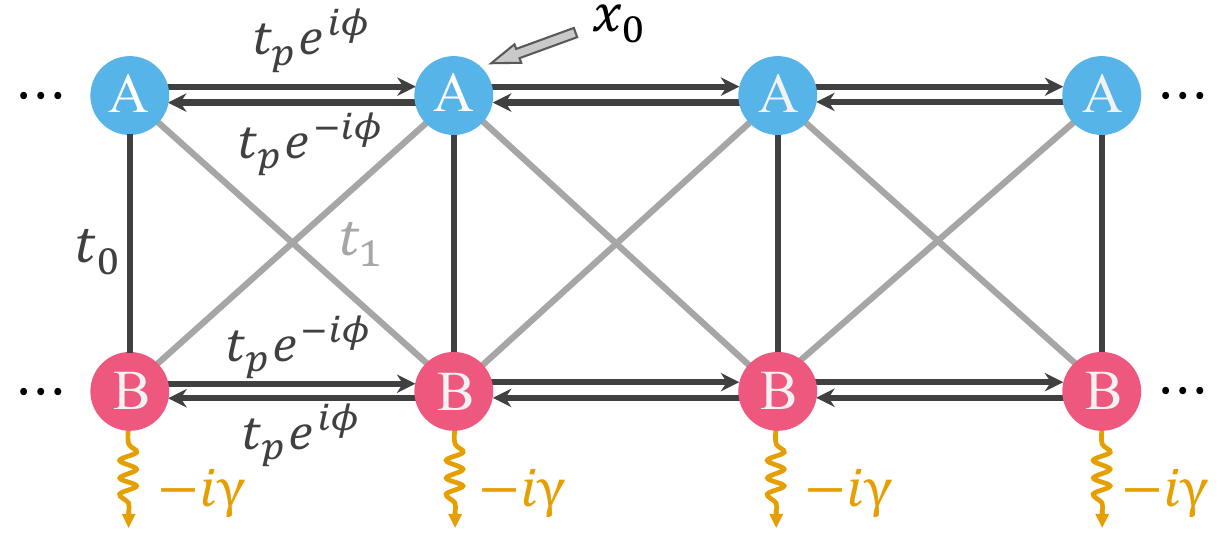}
  \caption{ Schematic of a one-dimensional lossy ladder lattice with nontrivial point-gap topology, which corresponds to the Hamiltonian in Eq.~(\ref{Eq(17)}). A nonzero Peierls phase $\phi$ is introduced via the hopping term $t_p$, which breaks the time-reversal symmetry of the system. } 
  \label{Fig3}
\end{figure}

\begin{figure}[t]
    \includegraphics[width=\columnwidth]{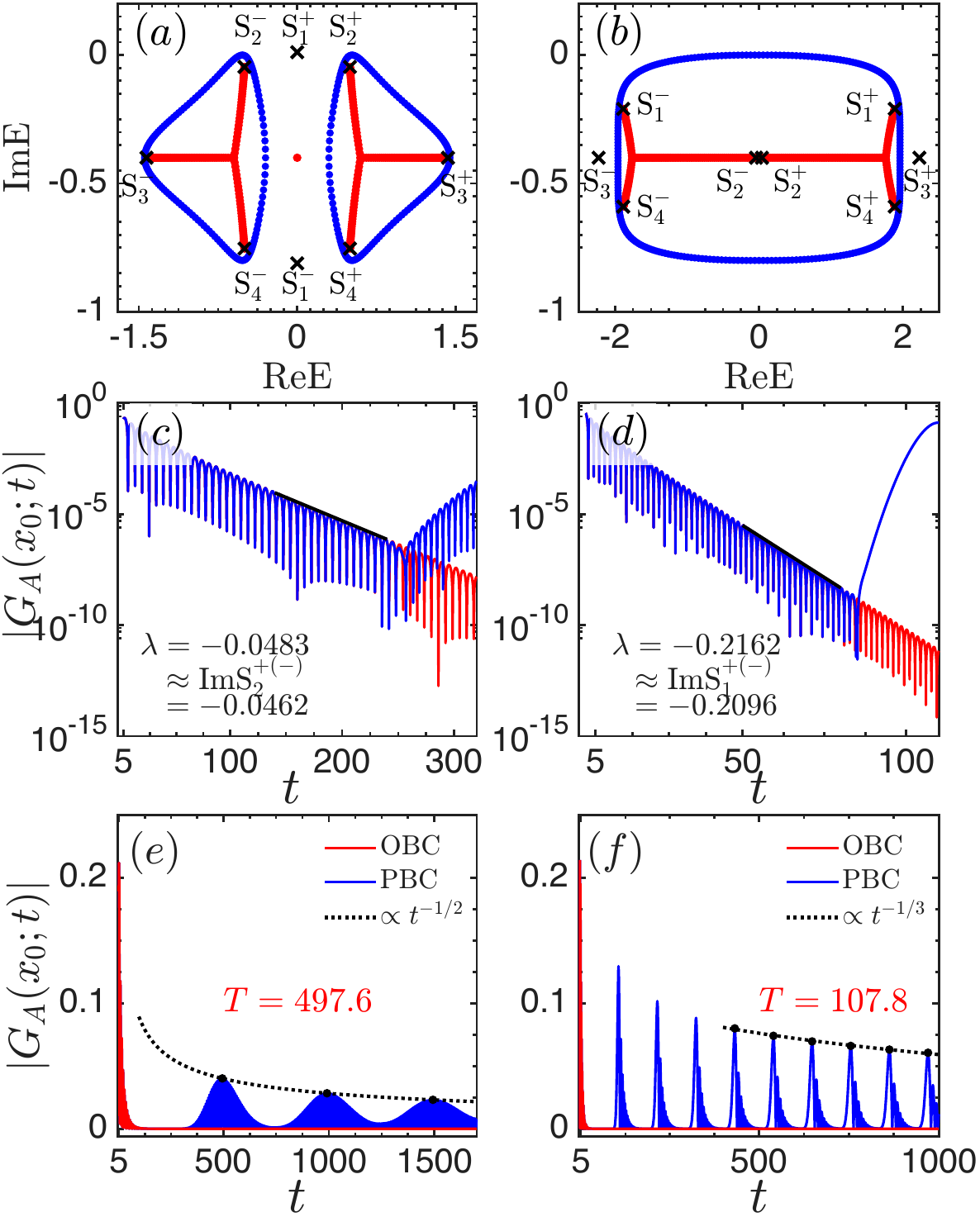}
  \caption{Energy spectra and corresponding
numerical results of $G_{A}\left(  x_{0};t\right)$ in an imaginary gap-closing system with nontrivial point-gap topology. (a)-(b) OBC (red points) and PBC (blue points) energy spectra $E_{\pm}$. Saddle points $S^{\pm}_i$ are explicitly marked by cross symbols. With the exception of $\operatorname{Im} S^{+}_1=0.01 >0 $ in (a), the imaginary parts of all other saddle points are negative. (c)-(d) Numerical results
for short-time evolution of $|G_{A}\left(  x_{0};t\right)|$. The fitted values  $\lambda$ are in agreement with the theoretical saddle-point predictions. (e)-(f) Numerical results for long-time evolution of $|G_{A}\left(  x_{0};t\right)|$  exhibit a power law envelope, fitted by black dashed line. The local peaks within this envelope recur with a period $T$. In (a), (c), and (e), the parameters are $t_0=0.5$, $t_1=0.5$, $t_p=0.3$, and $\gamma=0.8$; in (b), (d), and (f), the parameters are $t_0=1.0$, $t_1=0.5$, $t_p=0.7$, and $\gamma=0.8$. For all panels, the Peierls phase is $\phi =\pi/2$  and the number of unit cells is $L=150$. } 
  \label{fig4}
\end{figure}

We now consider imaginary gap-closing systems with a nontrivial point-gap
topology. As a concrete example, the model discussed here is also a two-band
model (see Fig.~\ref{Fig3}). The tight-binding Hamiltonian reads:
\begin{equation}
\begin{aligned}
H_{1}=& H+ \sum_{n} \left( t_{p}e^{i\phi}c_{n+1,A}^{\dagger}c_{n,A}%
+t_{p}e^{-i\phi}c_{n,A}^{\dagger}c_{n+1,A}
\right. \\
&\left. +t_{p}e^{-i\phi}c_{n+1,B}^{\dagger}c_{n,B}+t_{p}e^{i\phi}c_{n,B}^{\dagger}c_{n+1,B} \right),
\label{Eq(17)}
\end{aligned}
\end{equation}
where the hopping term $t_{p}$ between sites on the same sublattice carries a
Peierls phase, generating fluxes $\phi$ through the triangular plaquettes.
After applying a Fourier transformation and substitution $\beta\rightarrow e^{ik}$, we obtain
the generalized non-Hermitian Bloch Hamiltonian %
\begin{equation}
h_{1}\left(  \beta\right)  =h_{x}\left(  \beta\right)  \sigma_{x}+h_{z}\left(
\beta\right)  \sigma_{z}+h_{0}\left(  \beta\right)  I, %
\label{eq18}
\end{equation}
with 
\begin{equation*}
\begin{aligned}
h_{x}(\beta) &= t_{0} + t_{1}(\beta + \beta^{-1}), \\
h_{z}(\beta) &= it_{p}\sin\phi(\beta - \beta^{-1}) + i\gamma/2, \\
h_{0}(\beta) &= t_{p}(\beta + \beta^{-1})\cos\phi - i\gamma/2.
\end{aligned}
\end{equation*}
The spectrum corresponding to this
Hamiltonian is%
\begin{equation}
E_{1,\pm}\left(  \beta\right)  =h_{0}\left(  \beta\right)  \pm\sqrt{h_{x}%
^{2}\left(  \beta\right)  +h_{z}^{2}\left(  \beta\right)  }.
\end{equation}
Due to nonzero fluxes
$\phi$, this system breaks the time-reversal symmetry. Intuitively, the fluxes $\phi$ within the triangular plaquettes induce rotational motions, causing the $A$ and $B$ chains to favor propagation in opposite directions. The loss term then generates a net chiral motion along the $A$ chain by suppressing the backflow on the $B$ chain. Hence, onsite dissipation $\gamma$ induces NHSE under OBC \cite{PhysRevLett.125.186802}. As a result,
the PBC spectrum exhibits a nontrivial point-gap topology and differs
dramatically from its OBC counterpart. This is in sharp contrast to the
trivial point-gap model in Sec.~\ref{sec3}. To elucidate this
distinction and facilitate a clear discussion, we set $\phi=\pi/2$ for the
remainder of this section. As shown in Figs.~\ref{fig4}(a) and \ref{fig4}(b), two typical spectral configurations
emerge: the PBC spectra form either one or two closed loops, while the OBC
spectra consist of two open arcs enclosed by these PBC loops. 

The imaginary gap-closing condition of this system remains  $h_{x}\left(  \beta\right)  =0$ \cite{PhysRevB.109.214311}, that is, $\left\vert
t_{0}\right\vert \leq\left\vert 2t_{1}\right\vert $. However, these imaginary
gap-closing points are generally not saddle points of the corresponding nontrivial
point-gap system. For $\phi=\pi/2$, the saddle points defined by
$\frac{dE_{1,\pm}\left(  \beta\right)  }{d\beta}|_{\beta=\beta_{s}}=0$ satisfy
the following equation:
\begin{equation}
    h_{x}\left(  \beta_{s}\right)  h_{x}^{\prime}\left(  \beta_{s}\right)
+h_{z}\left(  \beta_{s}\right)  h_{z}^{\prime}\left(  \beta_{s}\right)  =0.
\end{equation}
For nonzero $t_{0}$, the solutions of this equation do not lie in the unit
circle $\left\vert \beta\right\vert =1$, implying that the corresponding
energies are not part of the PBC spectrum. In contrast, the endpoints of the
OBC spectra are attained for values of $\beta$ on the GBZ that correspond to
the saddle points of $E_{1,\pm}\left(  \beta\right)  $, as shown in Figs.~\ref{fig4}(a) and \ref{fig4}(b). For a general model, not all saddle points are located on the GBZ \cite{xue2025non,llbb-pcgk}. In fact, the corresponding characteristic equation $\det \left[ h\left( \beta \right) -E\right] =0$ gives an exact relation between $E$ and $\beta$. The saddle points are independent of the boundary conditions and are uniquely determined by the equation $\partial_{\beta}E(\beta)=0$, indicating that they correspond to double roots of the characteristic equation. Hence, these saddle points typically reside on the auxiliary generalized Brillouin zone (aGBZ) \cite{PhysRevLett.125.226402, PhysRevB.110.L140302,PhysRevB.111.165407}.

The dynamics of such a nontrivial point-gap model also
exhibit distinct features compared to the trivial point-gap 
model discussed in Sec.~\ref{sec3}. Here, we still consider the time evolution of the system under the initial conditions $\psi
_{x}^{A}\left( 0\right) =\delta _{x,x_{0}}$ and $\psi _{x}^{B}\left(
0\right) =0$. The local Green's function $G_{A}\left(  x_{0};t\right) $ exhibits
different scaling behavior on two distinct time scales. In the
short-time regime, the wave packet propagates within a finite range of the
lattice and does not traverse the entire system. Thus, as evidenced in Figs.~\ref{fig4}(c) and \ref{fig4}(d), the dynamics of the local Green's function $G_{A}\left(  x_{0};t\right)  $ are identical for both PBC and OBC. This equivalence allows for the
treatment of the system as an infinitely extended lattice, where the
discrete momentum $k$ can be approximated by a continuous variable. It
follows that $G_{A}\left(  x_{0};t\right) $ admits an integral representation:
\begin{equation}
  G_{A}\left(  x_{0};t\right) =\sum_{n=\pm }{\displaystyle\oint\nolimits_{\left\vert \beta\right\vert =1}}\frac{d\beta }{2\pi i\beta }g_{1,n}^{AA}\left( \beta
\right) e^{-iE_{1,n}\left( \beta \right) t}, 
\label{eq22}
\end{equation}
where $g_{1,n}^{AA}\left( \beta \right) =\left\langle A|R_{1,n}\left( \beta
\right) \right\rangle \left\langle L_{1,n}\left( \beta \right)
|A\right\rangle $, with $\left\vert R_{1,n}\left( \beta \right)
\right\rangle $ and $\left\langle L_{1,n}\left( \beta \right) \right\vert $
denoting the right and left eigenstates of the Bloch Hamiltonian (\ref{eq18}),
respectively. The asymptotic form of Eq. (\ref{eq22}) can also be evaluated using
the steepest descent method. Applying the complex Morse lemma \cite{milnor1963morse,matsumoto2002introduction},
the integral can be simplified by retaining the leading-order contributions
from the neighborhood of the saddle point $\beta _{s}$, yielding:%
\begin{equation}
 G_{A}\left(  x_{0};t\right)  \approx \sum_{n=\pm }\frac{%
g_{1,n}^{AA}\left( \beta _{s}\right) }{2\pi i\beta _{s}}e^{-iE_{1,n}\left(
\beta _{s}\right) t}\sim e^{\func{Im}\left( S\right) t},
\end{equation}where $S$ is the dominant saddle point energy. Since the imaginary gap-closing point is not a saddle point, the selection of the appropriate saddle
point through which the deformed contour $\mathcal{C}^{\prime }$ passes is
subtle. One might naively identify the dominant saddle point as the one
whose energy has the largest imaginary part. However, this conjecture is not
always valid. For instance, as shown in Figs.~\ref{fig4}(a) and \ref{fig4}(c), the saddle point energy $S_{1}^{+}$ has the largest imaginary part ($\func{Im}%
S_{1}^{+}=0.0111>0$) but is clearly not the dominant saddle point. Instead,
the short-time dynamics is governed by $S_{2}^{+}$ and $S_{2}^{-}$, which
share an imaginary part of $\func{Im}S_{2}^{+}=\func{Im}S_{2}^{-}=-0.0462$.
This value is in reasonable agreement with the numerical decay rate, $%
\lambda =$ $-0.0483$, extracted from the exponential envelope.

 In fact, when multiple saddle points are present, the deformed integration
contour $\mathcal{C}^{\prime }$ for each energy band $E_{n}\left( \beta
\right) $ can be expressed as a linear combination of Lefschetz thimbles \cite{berry1991hyperasymptotics,howls1997hyperasymptotics,pham1983vanishing,witten2011analytic,PhysRevB.90.035134,kanazawa2015structure},
such that $\mathcal{C}^{\prime }=\sum_{\sigma }n_{\sigma }D_{\sigma }$.
Here, each Lefschetz thimble $D_{\sigma }$ is the steepest descent path of $%
\func{Im}E_{n}\left( \beta \right) $ originating from a corresponding saddle
point $\beta _{s,\sigma }$. According to Morse theory, the coefficients $%
n_{\sigma }$ are determined by the intersections between the deformed
contour $\mathcal{C}^{\prime }$ (which is homologous to the original contour 
$\left\vert \beta \right\vert =1$) and the steepest ascent path $A_{\sigma }$
originating from the same saddle point. In particular, if $A_{\sigma }$
intersects the integration domain $\left\vert \beta \right\vert =1$ an even
number of times, the intersections must occur from opposite directions,
resulting in a zero net contribution, and therefore $n_{\sigma }=0$.
Consequently, the dominant saddle point can be identified as the one with a nonzero coefficient $n_{\sigma }$, whose associated energy
possesses the largest imaginary part.

In the long-time regime, the wave packet delocalizes throughout the whole
system. The local Green's function $G_{A}\left(  x_{0};t\right) $ exhibits
distinct dynamical behaviors for different boundary conditions, and the
system can no longer be treated as an infinitely long lattice. The
saddle-point description thus becomes invalid, and the dynamics are instead
dominated by imaginary gap-closing points, leading to algebraic decay
behavior. Specifically, for $\left\vert t_{0}\right\vert <2\left\vert t_{1}\right\vert $
and $t_{0}\neq 0$, the numerical results show that $G_{A}\left(  x_{0};t\right) $ follows a $t^{-1/2}$ power-law envelope [see
Fig.~\ref{fig4}(e)]. In contrast, for $\left\vert t_{0}\right\vert =2\left\vert
t_{1}\right\vert $, the envelope of $G_{A}\left(  x_{0};t\right)  $
changes to a decay scaling of $t^{-1/3}$, as shown in Fig.~\ref{fig4}(f). Furthermore,
for both types of scaling laws, the numerical results indicate that the peaks of $G_{A}\left(  x_{0};t\right)  $ recur with a period $T$.
These peaks are defined as the local maxima that conform to the envelope
scaling. This periodicity is closely related to the imaginary gap-closing
points, as will be discussed below.

For a general dissipative complex Bloch energy band $E_{n}\left( k\right) $,
the imaginary part determines the attenuation of the wavefunction amplitude,
while the real part is conventionally interpreted as a generalized energy.
Accordingly, we define the group velocity $v_{n,g}(k)$, as the derivative of
the real part of the energy:%
\begin{equation}
v_{n,g}\left( k\right) =\frac{d}{dk}\func{Re}E_{n}\left( k\right),\,\,k\in
\left( 0,2\pi \right]. 
\end{equation}%
This group velocity can be regarded as the characteristic speed of
information propagation, and the absolute value of its maximum can be
interpreted as the Lieb-Robinson bound for this non-Hermitian system \cite{lieb1972finite,PhysRevX.8.031079}.
Therefore, the aforementioned steepest descent method is valid when the
evolution time satisfies $t<L/\left( \left\vert v_{n,+}\right\vert
+\left\vert v_{n,-}\right\vert \right) $ , where $v_{n,+}=\max \left[ v_{n,g}%
\right] $ and $v_{n,-}$ $=\min [v_{n,g}]$ denote the maximum rightward and
leftward propagation velocities, respectively. When the imaginary gap-closing condition is satisfied, eigenmodes with energy near these closing
points decay the slowest and consequently govern the long-time dynamics. Therefore, we
consider the group velocity at these imaginary gap-closing points,
defined by $k_{0}$ where $\func{Im}E_{n}\left( k_{0}\right) =0$, yielding $%
v_{n,g}\left( k_{0}\right) $. The time required for a particle, initially
localized at a non-decaying position $x_{0}$, to traverse a circle and
return to $x_{0}$ via this imaginary gap-closing mode is $%
T=L/\left\vert v_{n,g}\left( k_{0}\right) \right\vert $. For the parameters
selected in Figs. 4(e) and 4(f), the theoretical predictions show that $T=500
$ and $T=107.1$, which are in reasonable agreement with the numerical
results, $T=497.6$ and $T=107.8$, respectively.

\begin{figure}[t]
    \includegraphics[width=\columnwidth]{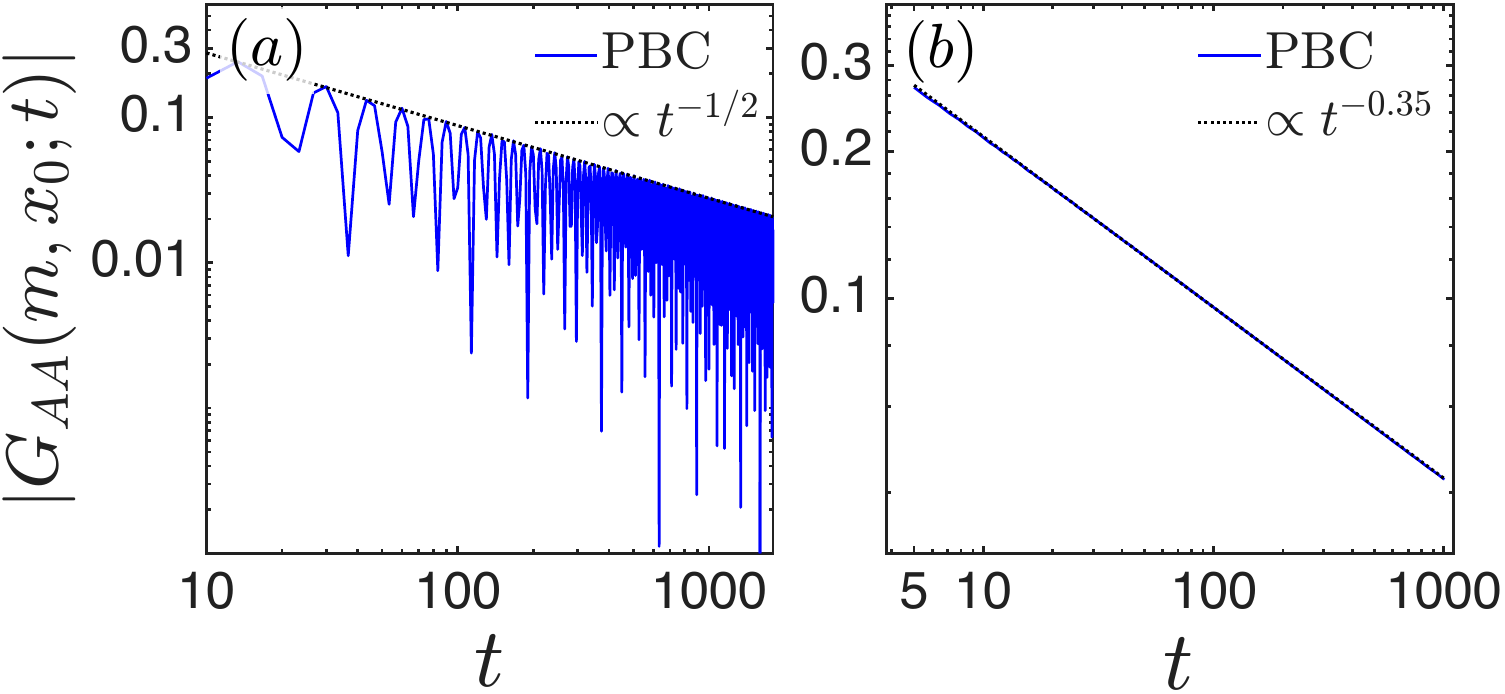}
  \caption{Numerical results for the amplitude of world-line Green's function $G_{AA}\left( m,x_{0};t\right) $ under PBC, where $m= x_0+v_{n,g}(k_0)t$. Here, $v_{n,g}(k_0)$ is the group velocity at the imaginary gap-closing point for the $E_n$ band.  The parameter values are: $t_0=0.5$, $t_1=0.5$, $t_p=0.3$, $\gamma=0.8$, and $v_{+,g}(k_0)=v_{-,g}(k_0)=-0.3$ in (a); $t_0=1.0$, $t_1=0.5$, $t_p=0.7$,  $\gamma=0.8$, and $v_{+,g}(k_0)=v_{-,g}(k_0)=-1.4$ in (b). For both panels, the initial conditions are $\psi_{x}^{A}(0)=\delta_{x,x_{0}}$ and $\psi_{x}^{B}(0)=0$, with the number of unit cells is $L$=150. } 
  \label{fig5}
\end{figure}

To elucidate the origin of the power law envelope scaling shown in Figs.~\ref{fig4}(e) and~\ref{fig4}(f), we analyze the temporal evolution of the amplitude $\psi
_{m}^{A}\left( t\right) $ along the space-time path $m=x_0+vt$, where $v$ is a
drift velocity. This quantity can be expressed in terms of the world-line
Green's function $G_{AA}\left( m,x_{0};t\right) $ \cite{PhysRevResearch.1.023013,xue2025non}, yielding: 
\begin{equation}
G_{AA}\left( m,x_{0};t\right) =\sum_{n=\pm }\int_{0}^{2\pi }\frac{dk}{2\pi }%
g_{1,n}^{AA}\left( k\right) e^{-i\left[ E_{n}\left( k\right) -kv\right] t}.
\end{equation}%
For convenience, momentum $k$ is used as the integration variable
instead of the Bloch phase factor $\beta =e^{ik}$. In the long-time limit,
the dynamics of $G_{AA}\left( m,x_{0};t\right) $ is determined by the saddle
points of the function $f_{n}\left( k\right) =E_{n}\left( k\right) -kv$.
Since $v$ is real, $f_{n}\left( k\right) $ has the same imaginary part as $%
E_{n}\left( k\right) $, resulting in $\func{Im}$ $f_{n}\left( k\right) \leq 0
$ for $k\in \left( 0,2\pi \right] $. When the drift velocity $v$ equals the
group velocity at an imaginary gap-closing point $k_{0}$, i.e., $%
v=v_{n,g}\left( k_{0}\right) $, we can expand $f_{n}\left( k\right) $ around 
$k_{0}$ with $\func{Im}f_{n}\left( k_{0}\right) =0$. At this point, we have $%
(dE_{n} /dk)|_{k=k_{0}}=v_{n,g}\left( k_{0}\right) $, implying
that $k_{0}$ is a saddle point of $f_{n}\left( k\right) $. Specifically, for
the case where $\left\vert t_{0}\right\vert <\left\vert 2 t_{1}\right\vert $
and $t_{0}\neq 0$, we find $(df_{n}/dk)|_{k_{0}}=0$ and $%
(d^{2}f_{n}/dk^{2})|_{k_{0}}\neq 0$. This indicates that $%
k_{0}$ is a second-order saddle point of $f_{n}\left( k\right) $. Applying
the saddle-point approximation reveals that the world-line Green's function $G_{AA}\left( m,x_{0};t\right) $ exhibits a $t^{-1/2}$ power-law scaling at
large $t$. This theoretical result is confirmed by our numerical
simulations, as shown in Fig.~\ref{fig5}(a). In contrast, when $\left\vert
t_{0}\right\vert =\left\vert 2 t_{1}\right\vert $, we have $(df_{n} /dk)|_{k_{0}}=(d^{2}f_{n} /dk^{2})|_{k_{0}}=0$ and $%
(d^{3}f_{n} /dk^{3})|_{k_{0}}\neq 0$. Thus, the imaginary gap-closing point $k_{0}$ becomes a
third-order saddle point of $f_{n}\left( k\right) $. Using a higher-order
saddle-point approximation, we find that $G_{AA}\left( m,x_{0};t\right) $ now
decays as a $t^{-1/3}$ power-law scaling at large $t$, which agrees
well with the numerical results shown in Fig.~\ref{fig5}(b). Therefore, the power-law envelope of $G_{A}\left(x_{0};t\right)$ observed in Figs.~\ref{fig4}(e) and
\ref{fig4}(f) on long time scales can be understood as follows. Considering $m=qL$,
where  $q$ is a positive integer, the word-line Green's function $G_{AA}\left( m,x_{0};t\right)$ is equal to the local Green's function $G_{A}\left(x_{0};t\right)$, due to the periodic boundary conditions.
Hence, $G_{A}\left(x_{0};t\right)$ obeys the same power-law decay as 
$G_{AA}\left( m,x_{0};t\right)$ when the drift velocity $v$ equals $%
v_{n,g}\left( k_{0}\right) $. The corresponding total evolution time is $%
t=m/\left\vert v_{n,g}\left( k_{0}\right) \right\vert =qT$, with the period
$T=L/\left\vert v_{n,g}\left( k_{0}\right) \right\vert $.

The preceding discussion can be extended to a more general case. If the
imaginary gap-closing point $k_{0}$ is an $n^{\prime}$-th order saddle point of $%
f_{l}\left( k\right) $, such that $(d^{p}f_{l}
/dk^{p})|_{k_{0}}=0$ for $p=1,2,\cdots ,n^{\prime}-1$ and $(d^{n^{\prime}}f_{l}
/dk^{n^{\prime}})|_{k_{0}}\neq 0$, the time evolution of $G_{AA}\left( m,x_{0};t\right)$ can be evaluated using a higher-order saddle point approximation.
Applying this method yields the asymptotic expression:%
\begin{equation}
    G_{AA}\left( m,x_{0};t\right) \sim \frac{%
  g_{1,n}^{AA}\left( k_{0}\right)  \Gamma \left( 1/n^{\prime}\right) }{ n^{\prime} \left[ -f_{l}^{\left( n^{\prime}\right) }\left(
k_{0}\right) /(n^{\prime}!)\right] ^{1/n^{\prime}}}  t^{-1/n^{\prime}}\ .
\end{equation}%
This result implies that $G_{AA}\left( m,x_{0};t\right)  $ exhibits a
characteristic scaling $t^{-1/n^{\prime}}$ . Consequently, the local Green's function $G_{A}\left(x_{0};t\right) $ obeys the same
power-law envelope scaling.

At the end of this section, we briefly examine the imaginary gap closing of the OBC spectrum. In a general non-Hermitian system, a point-gap nontrivial PBC $E_{PBC}$ spectrum encloses the corresponding OBC spectrum $E_{OBC}$, resulting in the relation $\max \left[\operatorname{Im} E_{OBC} \right]\leq \max \left[\operatorname{Im} E_{PBC}\right]$, where equality generally cannot be attained \cite{PhysRevLett.124.086801}. Consequently, for dissipative systems featuring the NHSE, the OBC spectrum typically does not satisfy the imaginary gap-closing condition. To enable the OBC spectrum of the Hamiltonian (\ref{Eq(17)}) to satisfy $\max \left[\operatorname{Im} E_{OBC} \right] = 0$, additional gain can be introduced to the entire system. This shifts the endpoints of the OBC spectrum onto the $\operatorname{Re} E$ axis. In this case, the saddle-point method remains applicable to the OBC system. However, the scaling behavior of the local Green’s function becomes dependent on the site $x_0$: $G_{A}\left(  x_{0};t\right)$ exhibits a $t^{-1/2}$ scaling in the bulk but a $t^{-3/2}$ scaling at the edge \cite{llbb-pcgk}.

\section{summary and Discussion}\label{sec5}

In summary, this work investigates the time evolution of quantum particles in
dissipative systems with imaginary gap closing. We find that the long-time
dynamics are governed by both imaginary gap-closing points and saddle points.
For systems with a trivial point gap, these two types of points can coincide,
resulting in a single scaling behavior characterized by an asymptotic
power-law decay of the local Green's function, proportional to
$t^{-1/n  }$, where $n$ is the order of the saddle point. In
contrast, for systems with a nontrivial point gap, the imaginary gap-closing
points generally do not coincide with the saddle points, leading to two
distinct dynamical regimes. In the short-time regime, the evolution is
dominated by the relevant saddle points and exhibits an asymptotic exponential
decay. In the long-time regime, the dynamics are controlled by the imaginary
gap-closing points, which also act as saddle points of the world-line Green's
function, resulting in an asymptotic power-law decay envelope scaling as
$t^{-1/ n^{\prime} }$. These findings are supported by both theoretical analysis, via the
saddle-point approximation method, and numerical simulations of concrete
models. Furthermore, the framework established here can be extended to other
non-Hermitian systems and may be experimentally verified on platforms such as
active mechanical metamaterials, electrical circuits, optical systems, and
cold atom setups.

\begin{acknowledgments}
The authors thank Xinyuan Gao for  helpful discussion. 
This work is supported by the Quantum Science and Technology-National Science and Technology Major Project (2024ZD0300600).
We acknowledge financial support
from the National Natural Science Foundation of China under Grant No. 92565105 and 12204395, Hong Kong RGC No. 14301425, No. 24308323, and No. C4050-23GF, Guangdong Provincial Quantum Science Strategic Initiative GDZX2404004, GDZX2505005, the Space Application System of China Manned Space
Program, and CUHK Direct Grant. Jinghui Pi also acknowledges support from the Postdoctoral Fellowship Program of CPSF under Grant No. 2025M773387.

\end{acknowledgments}

\appendix

\section{Saddle points of the analytical function }\label{appA}

In this appendix, we discuss the properties of saddle points in the complex
plane. For an analytic function $f(z)$ with complex variable $z=x+iy$, we
consider the points where the derivative is zero, that is, $df(z)/dz=0$. In the
following, we show that such points are always saddle points. By denoting
$f\left(  z\right)  =u\left(  x,y\right)  +iv\left(  x,y\right)  $, it follows
that $u\left(  x,y\right)  $ and $v\left(  x,y\right)  $ satisfy the
celebrated Cauchy-Riemann equations:%
\begin{equation}
 \frac{\partial u}{\partial x}=\frac{\partial v}{\partial y}, \qquad  \frac{\partial
u}{\partial y}=-\frac{\partial v}{\partial x}.  
\label{A1}
\end{equation}
As a result, the real part $u$ and the imaginary part $v$ are both harmonic
functions, which satisfy Laplace's equations:%
\begin{equation}
\frac{\partial^{2}u}{\partial x^{2}}+\frac{\partial^{2}u}{\partial y^{2}%
}=0, \qquad  \frac{\partial^{2}v}{\partial x^{2}}+\frac{\partial^{2}v}{\partial y^{2}%
}=0.
\end{equation}
These equations indicate that the second derivatives in the $x$ and $y$
directions are always opposite. Therefore, at a critical point $z_{0}$ where
$df(z)/dz=0$, $z_0$ is a saddle point for both $u$ and $v$. For example, if $u\left(  z_{0}\right)  $ is a local
maximum in the $x$ direction (i.e.,$\frac{\partial^{2}u}{\partial x^{2}}<0$),
then $u\left(  z_{0}\right)  $ is also a local minimum in the $y$ direction
(i.e., $\frac{\partial^{2}u}{\partial y^{2}}>0$).

Furthermore, the saddle point can be related to the steepest descent/ascent
lines of $v\left(  x,y\right)  $. The gradient of $v(x,y)$ corresponds to the
vector field $(\partial v/\partial x,\partial v/\partial y)$, and the
streamlines of this vector field are determined by:%
\begin{equation}
 \frac{dx}{\left(  \partial v/\partial x\right)  }=\frac{dy}{\left(  \partial
v/\partial y\right)  }.   
\end{equation}
On the other hand, condition $\operatorname{Re}f\left(  z\right)
=\operatorname{Re}f\left(  z_{0}\right)  $ defines several curves along which
the total derivative of $u\left(  x,y\right)  $ is zero. Thus, these curves
satisfy the following equation:
\begin{equation}
\frac{dy}{dx}=-\frac{\left(  \partial u/\partial x\right)  }{\left(  \partial
u/\partial y\right)  }.
\end{equation}
According to the Cauchy-Riemann equations (\ref{A1}), these curves coincide with the
streamlines of the vector field $\nabla v(x,y)$. Therefore, these curves
follow either the direction of the steepest ascent of $v(x,y)$ or the direction of the
steepest descent. For a given saddle point $z_{0}$, the conditions that (i)
$u\left(  x,y\right)  $ is stationary and (ii) $\operatorname{Im}$ $v(x,y)$ is
maximized at the point $z_{0}\in J$ imply that the choice of $J$ is unique. It
comprises the two streamlines that originate from $z_{0}$ and follow the
direction of steepest change in $v(x,y)$.

\section{Saddle point approximation method}\label{appB}
In this appendix, we provide a detailed discussion of the saddle-point (or
steepest-descent) method. The central aim of this method is to calculate the
asymptotic expression for integrals of the form $\int g\left(  z\right)
e^{\lambda f\left(  z\right)  }dz$ in the limit of a large parameter $\lambda$.

When the integration variable is a real number, such an integral can be
evaluated using Laplace's method. Specifically, for a real integral $I\left(
\lambda\right)  =\int_{a}^{b}g\left(  x\right)  e^{\lambda f\left(  x\right)
}dx$, if $f\left(  x\right)  $ attains its unique maximum within the interval
at a point $c\in\left(  a,b\right)  $, then as $\lambda\rightarrow\infty$, the
integral has the asymptotic form:%
\begin{equation}
I\left(  \lambda\right)  \sim\sqrt{\frac{2\pi}{-\lambda f^{\prime\prime
}\left(  c\right)  }}g\left(  c\right)  e^{\lambda f\left(  c\right)  }.
\end{equation}
This expression holds provided that $f^{\prime\prime}\left(  c\right)  \neq0$
and $g\left(  c\right)  \neq0$. As the integrand is sharply peaked around $c$
for large $\lambda$, one can approximate $f(x)$ by its Taylor expansion around
$c$ up to the second order and perform a Gaussian integral to obtain the
dominant part of the integral. If $f(x)$ has more than one local maximum,
correction terms for the asymptotic expression arise at large but finite
$\lambda$. One can perform a similar Gaussian-integral approximation around
each individual local maximum $c_{i}$, leading to:
\begin{equation}
I\left(  \lambda\right)  \sim\sum_{i}\sqrt{\frac{2\pi}{-\lambda f^{\prime
\prime}\left(  c_{i}\right)  }}g\left(  c_{i}\right)  e^{\lambda f\left(
c_{i}\right)  }.
\end{equation}

When the integration is a contour integral, such as $%
{\displaystyle\oint\nolimits_{\mathcal{C}}}
g\left(  z\right)  e^{\lambda f\left(  z\right)  }dz$ in the complex $z$-plane, we
parametrize the contour $\mathcal{C}$ with $z=\gamma\left(  x\right)  $. In
this case, the dominant part of the integral is determined by the maxima of
$\operatorname{Re}f\left(  x\right)  $, and we can still employ Laplace's
approximation. As $\operatorname{Re}f\left(  x\right)  $ reaches its maximum
value at $x=c$, in the neighborhood of $c$, we have:%
\begin{equation}
f\left(  x\right)  \approx f\left(  c\right)  +i\lambda\left(  x-c\right)
\operatorname{Im}f^{\prime}\left(  c\right)  +\frac{\lambda\left(  x-c\right)
^{2}}{2}f^{\prime\prime}\left(  c\right)  .
\end{equation}
Therefore, the additional factor $e^{i\lambda\left(  x-c\right)
\operatorname{Im}f^{\prime}\left(  c\right)  }$ oscillates strongly for large
$\lambda $, making the Gaussian integral indeterminate. However, by Cauchy's
theorem, we can deform the contour $\mathcal{C}$ into a new contour
$\mathcal{C}^{\prime}$ without altering the integral's value. To facilitate
the calculation, the new contour $\mathcal{C}^{\prime}$ is chosen such that
$\operatorname{Im}f\left(  z\right)  $ is constant along the path
$\mathcal{C}^{\prime}$. In addition, there is a point $z_{s}\in\mathcal{C}%
^{\prime}$ where $\operatorname{Re}f\left(  z\right)  $ attains its maximal
value on $\mathcal{C}^{\prime}$. Under these two conditions, we can safely use 
Laplace's approximation and obtain the asymptotic expression:%
\begin{equation}
I\left(  \lambda\right)  \sim\sqrt{\frac{2\pi}{-\lambda f^{\prime\prime
}\left(  z_{s}\right)  }}g\left(  z_{s}\right)  e^{\lambda f\left(
z_{s}\right)  }.
\label{B4}
\end{equation}
This is valid provided that $f^{\prime\prime}\left(  z_{s}\right)  \neq0$ and
$g\left(  z_{s}\right)  \neq0$. By denoting $z_{s}=\gamma\left(  c\right)  $,
it is straightforward to show that $\frac{d}{dx}f\left(  \gamma\left(
c\right)  \right)  =\gamma^{\prime}\left(  c\right)  f^{\prime}\left(
z_{s}\right)  =0$. Since $\gamma^{\prime}\left(  x\right)  $ should be
non-vanishing for non-singular parametrization, it follows that $f^{\prime
}\left(  z_{s}\right)  =0$. As a result, $z_{s}$ is a saddle point of
$f\left(  z\right)  $. The integral (\ref{B4}) is asymptotically dominated by the neighborhood of $z_{s}$, similar to the real case. This suggests that other
parts of the contour of integration are less relevant, and the condition on
the contour $\mathcal{C}^{\prime}$ can be relaxed. Specifically, the
requirement that $\operatorname{Im}f$ remains constant along the entire
contour $\mathcal{C}^{\prime}$ can be relaxed to it being stationary near
$z_{s}$.

\nocite{*}

\bibliography{apssamp}

\end{document}